\documentclass[singlecolumn,floatfix,amsmath,amssymb,osajnl2,showpacs]{revtex4}

\usepackage{graphicx}
\usepackage{color}
\usepackage{amssymb}
\usepackage{amsmath}
\usepackage{braket}
\usepackage{verbatim}

\begin{document}

\author{Karthik Sasihithlu}%
\email{k.sasihithlu@imperial.ac.uk}
\affiliation{%
Blackett Laboratory, Imperial College\\
London, UK, SW7 2AZ
}%
\author{Girish Agarwal}
\affiliation{%
Institute for Quantum Science and Engineering,  \\ Department of Biological and Agricultural Engineering, Department of Physics and Astronomy \\
Texas A\&M University, TAMU4242\\
College Station, Texas 77843, USA,
}%

\date{\today}



\title{Dynamic near-field heat transfer between macroscopic bodies for nanometric gaps}

\begin{abstract}
The dynamic heat transfer between two half-spaces separated by a vacuum gap due to coupling of their surface modes is modelled using the theory that describes the dynamic energy transfer between two coupled harmonic oscillators each separately connected to a heat bath and with the heat baths maintained at different temperatures. The theory is applied for the case when the two surfaces are made up of a polar crystal which supports surface polaritons that can be excited at room temperature and the predicted heat transfer is compared with the steady state heat transfer value calculated from standard fluctuational electrodynamics theory. It is observed that  for small time intervals the value of heat flux can be significantly higher than that of steady state value.
\end{abstract}

\maketitle

The theory of photon mediated heat transfer between macroscopic objects in close proximity to each other 
and separated by a vacuum gap has traditionally been treated using the macroscopic Rytov's fluctuational electrodynamics theory which assumes local thermodynamic equilibrium in the bodies in question \cite{polder71, loomis94, pendry1999radiative, joulain05a}.
This heat transfer comprises of contributions from both long-range radiative modes as well as near-field evanescent and surface modes \cite{joulain05a, sasihithlu2011proximity}. When thermally excited, contributions from surface modes - which are electromagnetic eigenmodes of the surface and are characterized by their field decaying exponentially on either side of the interface - dominate the heat transfer between the surfaces when the gap is less than the thermal wavelength of radiation. This is primarily due to a peak in the density of electromagnetic states at such frequencies where these surface modes are resonantly activated  as evidenced from the dispersion relation for these modes \cite{joulain05a}. In particular, for this effect to be prominent at room temperature the surfaces should be made up of a polar crystal such as silicon carbide or silica which supports surface-phonon polariton modes in the infra-red wavelength around 10 $\mu$m and can thus be thermally excited at these temperatures.

 In general, resonant excitation of surface modes plays an important role in several phenomena and applications including: decreased lifetime of molecules close to metal surfaces \cite{chance1978molecular},  surface enhanced raman spectroscopy \cite{le2008principles}, thermal near-field spectroscopy \cite{jones2012thermal, babuty2013blackbody},  concept of perfect lens \cite{Pendry00a} and thermal rectification \cite{otey2010thermal}. The study of coupling of surface modes across surfaces is significant as it not only plays an important role in heat transfer but also in the van der Waals and Casimir force between them \cite{van1968macroscopic, sernelius2011surface}. A coupled harmonic oscillator description for the heat transfer between nanoparticles due to coupling of surface modes was arrived at by Biehs and Agarwal \cite{biehs2013dynamical} and estimates for both dynamic and steady state heat transfer values were arrived at.  
 Barton \cite{barton2015classical} has considered the heat flow between two harmonic oscillators using Langevin dynamics and has extended  this model to planar surfaces but has limited his description to the steady-state heat flow.
 %
Yu et. al., \cite{yu2017ultrafast} have recently analysed the dynamics of radiative heat exchange between graphene nanostructures using fluctuational electrodynamics principles and have observed thermalization within femtosecond timescales. This ultrafast heat exchange due to the time varying temperatures of the nanostructures is a resultant of the low heat capacity of the graphene nanostructures and the coupling of the large plasmonic fields.  A similar analysis for the radiative heat transfer between graphene and a hyperbolic material  has been carried out by Principi et. al., \cite{principi2017super} where they observe thermalization in picosecond timescales.
   In this paper we model the dynamic heat transfer contribution from coupling of surface modes across two dielectric planar surfaces  using the master equation description of two coupled-harmonic oscillators interacting with their respective heat baths and compare the steady-state results with those obtained from fluctuational electrodynamics principles available in literature \cite{loomis94, joulain05a}. Our work differs from that in Ref. \cite{yu2017ultrafast, principi2017super} in that we are interested in analysing how the coupling between the surface modes and the resultant heat transfer between the two surfaces, which are maintained at fixed temperatures, relaxes to steady state.
 \begin{figure}[h!]
\includegraphics[scale=0.55]{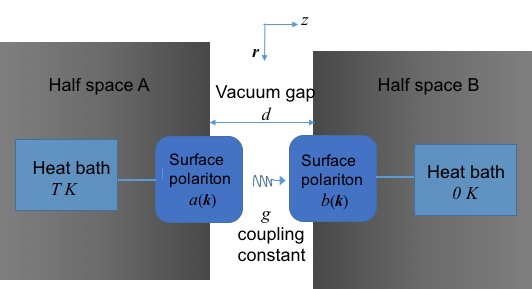}
\caption{Coupling between two surface modes each connected to its own heat bath that is analysed in this work. The two half-spaces are assumed to be made of the same material which is local and dispersive so that the complex dielectric function varies only with the frequency $\omega$. Only modes (denoted by the amplitudes $a(\mathbf{k})$ and $b(\mathbf{k})$) of the same in-plane wavevector component $\mathbf{k}$ couple across the vacuum gap $d$.  The coupling constant $g$ between the surface modes is derived using Maxwell's equations}
\label{FigConfig}
\end{figure}

The paper is arranged as follows: in Section \ref{sec1} the results of the dynamics of heat transfer between two coupled HO each in contact with a heat bath are summarised. While the theory has been described in detail in Ref. \cite{biehs2013dynamical}, for sake of completion the main results have been reproduced here with added details. Such a system has also been analysed previously in the context of analysing the dependence of mean interaction energy on the temperatures of heat baths \cite{dorofeyev2013coupled},  and entanglement between two particles \cite{ghesquiere2013entanglement}. In Section \ref{sec2} the theory of dynamics of heat transfer between two coupled HO is extended to that between two half spaces by analysing the coupling between two interacting planar surface modes using Maxwell's equations.  In Section \ref{section3} numerical values for the heat flux derived in Sec. \ref{sec1} and \ref{sec2} are plotted for the particular case of two silicon carbide half-spaces and compared with calculations from fluctuational electrodynamics principles. 

\section{Heat transfer dynamics between two harmonic oscillators}
\label{sec1}
A surface polariton located at the interface between a half-space located at $z>0$ and vacuum will have field of the form $\text{exp}(i \mathbf{k}.\mathbf{r} - i \omega t + i k_z z )$, where the in-plane component $\mathbf{k} = (k_x, k_y)$  and the $z$-component $k_z$ of the wavevector are related as $\mathbf{k}^2+k_z^2 = (\omega/c)^2$  with $c$ being the velocity of light,   $\omega$ the frequency of the planar wave in the vacuum gap of this system,  and $\mathbf{k}$ being limited by $|\mathbf{k}| > \omega/c$. %
The surface polariton exists for the ($\mathbf{k}, \omega$) pair that satisfies the well known dispersion relation \cite{maier2007plasmonics}  $|\mathbf{k}| = (\omega/c)\sqrt{ \varepsilon(\omega)/(\varepsilon(\omega)+1) }$ so that the frequency $\omega_0$ of the surface polariton is characterized by the wavevector $\mathbf{k}$.
 Thus one can characterize the surface excitations in terms of oscillators with frequency $\omega_0(\mathbf{k})$, complex amplitude $a(\mathbf{k})$ and half line width $\gamma(\mathbf{k})$ with both $\omega_0(\mathbf{k})$ and $ \gamma(\mathbf{k})$ determined from the dispersion relation.   

When two half-spaces are  far apart then each half-space consists of oscillators with frequencies $\omega_0(\mathbf{k})$  which are in thermal equilibrium at the temperature of the half-space. When they are  brought close to each other as shown in Fig. \ref{FigConfig} then the surface polariton modes of half-space A interact with the surface polariton modes of half-space B. However, it should be borne in mind that the wavevector $\mathbf{k}$ is conserved for fields across planar interfaces. This implies an effective coupling of the form:
\begin{equation}
H_\text{I}  = \sum_{\text{allowed $\mathbf{k}$ values }}  \hbar g(\mathbf{k}) \left[ a^\dagger(\mathbf{k}) b(\mathbf{k}) + b^\dagger(\mathbf{k}) a(\mathbf{k}) \right]
\label{intH}
\end{equation}
where $b(\mathbf{k})$ is the complex amplitude of the surface polariton on the half-space B, $g(\mathbf{k})$ is the coupling constant (units of rad s$^{-1}$) and we neglect non-resonant contributions to the coupling (i.e. we use the rotating wave approximation). Thus the mode $a(\mathbf{k})$ couples to the mode $b(\mathbf{k})$ only. There are no coupling terms of the form $a^\dagger(\mathbf{k}_1) b(\mathbf{k}_2)$, $\mathbf{k}_1 \neq \mathbf{k}_2$. Thus we can consider coupling between two oscillators $a(\mathbf{k})$ and  $b(\mathbf{k})$ for a fixed $\mathbf{k}$, and at the end of calculation sum over all modes. Note also that in the non-retarded limit the surface modes occur for all values of $\mathbf{k}$.
This coupling results in a heat flow due to interaction between the modes which is a dynamical process, and which can be described using the master equation approach. 
The structure of the weakly coupled master equation for two interacting oscillators (with amplitudes $a(\mathbf{k})$ and $b(\mathbf{k})$))  is well known  \cite{agarwal2013quantum} and is given by:
\begin{equation}
\begin{split}
\frac{\partial \rho_S}{\partial t} &= - i \omega_0[a^\dagger a, \rho_S] - i \omega_0[b^\dagger b, \rho_S] - i g [a^\dagger b + b^\dagger a, \rho_S] \\ & - \gamma ( n_1 +1) (a^ \dagger a \rho_S - 2 a \rho_S a^\dagger + \rho_S a^\dagger a) - \gamma  n_1  (a  a^\dagger \rho_S - 2 a^\dagger \rho_S a  + \rho_S a  a^\dagger) - \gamma  (b^ \dagger b \rho_S - 2 b \rho_S b^\dagger + \rho_S b^\dagger b);
\end{split}
\end{equation}
Here $\rho_S$ is the density matrix for the two oscillator system. The operators $a$, $a^\dagger$, $b$, $b^\dagger$ satisfy Boson algebra with the symbol $ [..]$ denoting the commutation operator, and for brevity we drop the argument $\mathbf{k}$. We have assumed that the two half-spaces have identical dielectric properties.  The half-space A is at temperature $T$ so that $n_1$ is the excitation of the mode $a$ i.e., $n_1 =  1/(\text{exp} \left(\hbar \omega_0/(k_B T)-1\right))$  where, $k_B$ is the Boltzmann constant. 
The half-space B is at zero temperature. We also take $g \ll \omega_0$ which is later shown to be a valid assumption for the system under consideration. Higher order terms in the master equation would be necessary when this assumption is not satisfied \cite{walls1970higher}. This master equation approach is valid only for oscillators where  $\gamma/\omega_0 \ll 1$, and can be used to describe dynamics for time scales $t \gg  1/\omega_0 $ \cite{carmichael2003statistical}.

 From this, and relating expectation value of an operator $\langle G \rangle$  to the density matrix $ \rho_S$ using the standard relation $\partial \langle G \rangle /\partial t   =  \text{Tr}(\partial \rho_S/\partial t\,\, G )$ we obtain the rate of change of the excitation of surface polariton in B to be:
\begin{equation}
\frac{\partial}{\partial t} \langle b^\dagger b \rangle = -i g \left(  \langle b^\dagger a \rangle - \langle a^\dagger b\rangle \right) - 2 \gamma  \langle b^\dagger b \rangle
\label{dyna1}
\end{equation}
The first term on the right hand side represents the change in the excitation of surface polariton in B due to the flow of thermal excitation from the surface polariton in A to that in B due to the temperature gradient. This enables us to identify  the heat transferred (in units W) to surface polariton in B as:
\begin{equation}
 \langle P \rangle (t) = -i g\hbar \omega_{0}  \left(  \langle b^\dagger a \rangle  -\langle a^\dagger b \rangle\right)
  \label{eqHO}
 \end{equation}
To find $\langle P \rangle (t)$ we follow the procedure used to obtain Eq. \ref{dyna1} to get the dynamic matrix equation:
\begin{equation}
\dot{{\bf x}} = \mathbb{A} {\bf x} + {\bf a}
\label{matrix}
\end{equation}
with  ${\bf x} = (\langle a^\dagger a \rangle , \langle b^\dagger b \rangle  , i (\langle b^\dagger a \rangle  - \langle a^\dagger b\rangle) )^T  $,  ${\bf a} = (2 \gamma n_1, 0, 0)^T  $ and
\begin{equation}
\mathbb{A} =   \begin{bmatrix} -2 \gamma & 0 &  g \\ 0 & -2 \gamma & -g \\ -2 g &  2g & -2 \gamma   \end{bmatrix} \nonumber
\end{equation}
 The value for  $i \left(  \langle b^\dagger a \rangle  -\langle a^\dagger b \rangle\right)$ in Eq. \ref{eqHO} from which all time dependent and steady state characteristics of the heat transfer can be obtained, is got by solving the dynamic matrix equation in Eq. \ref{matrix}.
with the initial conditions:  $\langle a^\dagger a \rangle|_{t=0} = n_1;\,\,\, \langle b^\dagger b \rangle|_{t=0} = 0;\,\,\,  i\langle b^\dagger a  - a^\dagger b \rangle |_{t=0} = 0 $, i.e.,  we consider the interaction to switch on at time $t=0$ and analyse the dynamical evolution of the system. We thus get the expression for heat transfer from Eq. \ref{eqHO} as:
\begin{equation}
\langle P \rangle (t) = \frac{\hbar \omega_0 g^2 \gamma n_1}{(g^2 + \gamma^2)} \Big[ 1 -  e^{- 2 t \gamma} \cos{2tg} - (\gamma/g)e^{- 2 t \gamma} \sin{2tg} \Big]+ \hbar \omega_0 g e^{-2t \gamma} \sin{2 g t} \label{analytical}
\end{equation}



 \section{Heat transfer between two half spaces}
 \label{sec2}

In this section we  discuss how the different parameters in the master equation can be obtained from the nature of surface polaritons for a system of two dielectric half spaces separated by a  vacuum gap and subsequently calculate the heat transfer for this system.  Consider first a system of two coupled oscillators with natural frequency $\omega_0$, half line-width $\gamma$, coupling constant $\xi \omega_0^2$, and displacements $x_1(\omega)$, $x_2(\omega)$ where $x_{1/2}(\omega) = \int e^{i \omega t} \, x_{1/2}(t) \, dt$ which are related as:
\begin{eqnarray}
\omega^2 x_1 (\omega) &=& \omega_0^2 x_1(\omega) - 2 i \omega \gamma x_1 (\omega) +  \xi \omega_0^2 x_2 (\omega) \\ \nonumber
\omega^2 x_2 (\omega) &=& \omega_0^2 x_2 (\omega) - 2 i \omega \gamma x_2 (\omega) + \xi \omega_0^2 x_1 (\omega)
\label{coupledho}
\end{eqnarray}
The eigenfrequencies for such a system are are given by:
\begin{equation}
\omega_{\pm}=\sqrt{\omega_{0}^2 (1 \pm \xi) - \gamma^2} - i \gamma;
\label{Eq1}
\end{equation}
To find the the equivalent coupling constant $\xi$  between surface modes  we  find the eigenmodes of the configuration of two flat surfaces
separated by a vacuum gap using Maxwell's equations and compare the expression with that of Eq. \ref{Eq1} for two  coupled harmonic oscillators. Consider two half-spaces separated by
a gap of width $d$ occupying the regions $z<-d/2$ and $z>d/2$ as shown in Fig. \ref{FigConfig}. 
From the dispersion relation for surface polaritons $\omega(\mathbf{k})$ and close to the surface polariton frequency the surface mode is seen to acquire an electrostatic character with $|\mathbf{k}| \rightarrow \infty$   \cite{maier2007plasmonics}.  
In this electrostatic limit, satisfying continuity of perpendicular component of displacement
field gives the condition for surface modes as:
\begin{eqnarray}
\varepsilon(\omega\pm)=\begin{cases}
\left(1-e^{|\mathbf{k}| d}\right)/\left(1+e^{|\mathbf{k}| d}\right)\\
\left(1+e^{|\mathbf{k}| d}\right)/ \left(1-e^{|\mathbf{k}| d}\right)
\end{cases}
\label{condition}
\end{eqnarray}
where
$\varepsilon(\omega)$ is the dielectric function of the half-space. Taking the Lorentz model for the dielectric function:
\begin{equation}
\varepsilon(\omega)=\varepsilon_\infty \left( 1+\dfrac{\omega_L^2 - \omega_T^2}{\omega_{T}^{2}-\omega^{2}-i\omega\Gamma}\right)
\label{dielectric}
\end{equation}
and solving for $\omega$ in Eq. \ref{condition} in the low-loss limit $\Gamma\rightarrow0$ and with some algebra we obtain the eigenfrequencies of the coupled surface modes to be of the form:
 \begin{equation}
 \omega(\pm) = \sqrt{\omega_0^{'2}( \mathbf{k}, d)\left( 1 \pm \xi'(\mathbf{k}, d) \right) - \frac{\Gamma^2}{4}} - i \frac{\Gamma}{2}
 \label{eigensurface}
  \end{equation}
  where,
  \begin{equation}
 \omega'_0(\mathbf{k}, d) = \sqrt{\dfrac{2 e^{-  |\mathbf{k}| d}  \left(\varepsilon_\infty(\omega_L^2 + \omega_T^2) \cosh(|\mathbf{k}| d) + (\varepsilon_\infty^2 \omega_L^2 + \omega_T^2) \sinh(|\mathbf{k}| d) \right)}{(\varepsilon_\infty + 1)^2 - (\varepsilon_\infty - 1)^2 e^{- 2 |\mathbf{k}| d}}}
 \label{omegap}
 \end{equation}
 and
 \begin{equation}
 \xi'(\mathbf{k}, d) = \dfrac{\varepsilon_\infty(\omega_L^2 - \omega_T^2)}{\varepsilon_\infty(\omega_L^2 + \omega_T^2) \cosh (|\mathbf{k}| d)+ (\varepsilon_\infty^2 \omega_L^2 + \omega_T^2)\sinh(|\mathbf{k}| d) } ;
 \label{xip}
 \end{equation}
In the limit of large gaps $|\mathbf{k}| \, d \rightarrow \infty$ $,\omega_0'(\mathbf{k},d) \approx \omega_0$ where $\omega_0$ is the surface polariton frequency for a single half-space given by $\omega_0 = \sqrt{(\varepsilon_\infty \omega_L^2 + \omega_T^2)/(\varepsilon_\infty+1)} $ and the coupling parameter $\xi'(\mathbf{k},d) \approx 0$.  For the case when $\varepsilon_\infty =1$ these expressions reduce to the simple forms: $\omega_0' = \sqrt{(\omega_L^2 + \omega_T^2)/2}$ and $\xi'(\mathbf{k}, d)  = e^{- |\mathbf{k}| d}(\omega_L^2 - \omega_T^2)/(\omega_L^2 + \omega_T^2) $ \cite{barton2015classical}. This method of finding the eigenmodes of the surface modes in the electrostatic limit is well known and has been previously employed, for example, in the context of deriving the van der Waals force \cite{van1968macroscopic} and the steady state van der Waals heat transfer \cite{barton2015classical}.
By comparing Eq. \ref{eigensurface} with Eq. \ref{Eq1} we can relate parameters of the harmonic oscillator model with those of the surface modes:
\begin{equation}
\omega_0 \rightarrow \omega_0'(\mathbf{k}, d); \,\,\,\xi \rightarrow \xi'(\mathbf{k}, d); \,\,\, \gamma\rightarrow \Gamma/2
\label{gamma}
\end{equation}
We can thus use the results from Sec. \ref{sec1} to find the heat transfer between the two flat surfaces due to coupling between the surface modes.  As noted earlier, we add the contributions from all the harmonic oscillator modes (labelled by $\mathbf {k}$)  observing that heat transfer does not result in change in momentum and that only surface modes of the same in-plane wave-vector component interact across the two half spaces. The heat flux between two half-spaces $P (t,d)$ (Wm$^{-2}$)  is then given by:
\begin{equation}
P(t,d) = \int_0^\infty \frac{ 1}{4 \pi^2} \,\langle P \rangle \, d^2 \mathbf {k}
\label{Q}
\end{equation}
where $\langle P \rangle (t, \mathbf{k},  d) $ is the rate of heat transfer (units of W) between two harmonic oscillators obtained from Eq. \ref{eqHO}. 
The relation between $g$ in Eq. \ref{eqHO} and $\xi$ in Eq. \ref{Eq1} can be derived by comparing the interaction Hamiltonians in the quantum mechanical picture and the classical model.  In the classical picture the potential energy of the Hamiltonian of the coupled harmonic oscillator system is given by:
\[
H_{\text{PE}}=\dfrac{1}{2}\omega_{0}^{2}x_{1}^{2}+\dfrac{1}{2}\omega_{0}^{2}x_{2}^{2}+\omega_{0}^{2}\xi x_{1}x_{2}
\]
where the interaction term is given from: $H_{\text{I}}=\omega_{0}^{2}\xi x_{1}x_{2}$.
The equivalent quantum mechanical model can be got by substituting the  displacements $x_{1}$and $x_{2}$ in terms of the commutator operators $a$,$a^\dagger$, $b$,$b^\dagger$ using the standard relations $x_1 = (a+a^\dagger)\sqrt{\hbar/(2\omega_0)}$, and $x_2 = (b+b^\dagger)\sqrt{\hbar/(2\omega_0)}$. Including only terms resulting in photon exchange we get:
\[
\hat{H}_{\text{I}}=\omega_{0}\xi(a^{\dagger}b+ab^{\dagger})\dfrac{\hbar}{2}
\]
Comparing with the interaction Hamiltonian in the quantum mechanical picture from Eq. \ref{intH} (for a particular $\mathbf{k}$ value)  gives us: $
g=\omega_{0}\xi/2$.  Substituting for the values of $\omega_0$ and $\xi$ from Eq. \ref{gamma} gives:
\begin{equation}
g(\mathbf{k}, d)=\dfrac{\omega_{0}'(\mathbf{k}, d) \xi'(\mathbf{k}, d)}{2}
\label{gvarywithgap}
\end{equation}

\section{Results}
\label{section3}

 \begin{figure}[h!]
\includegraphics[scale=0.40]{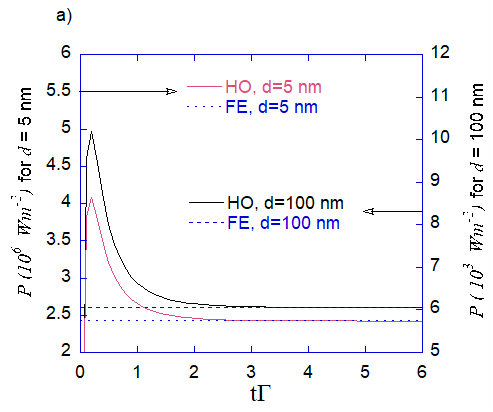}
\begin{center}$
\begin{array}{cc}
\includegraphics[scale=0.30]{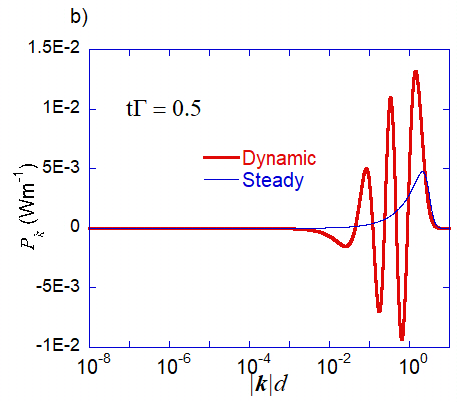}
\includegraphics[scale=0.30]{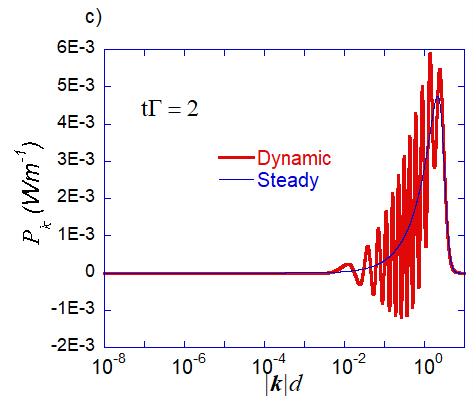}
\includegraphics[scale=0.30]{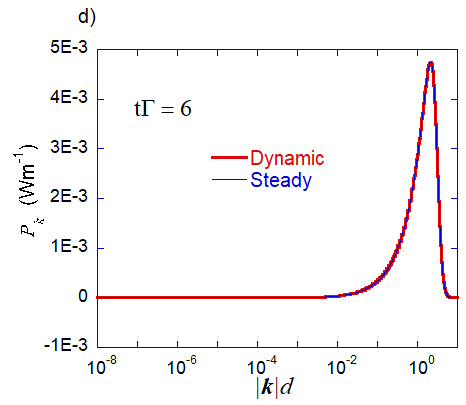}
\end{array}$
\end{center}
\caption{a) Plot of heat flux  at room temperature from Eq. \ref{Q} (indicated as `HO')  as a function of  nondimensional time $t \Gamma$ between two SiC half spaces for gaps $d=5$ nm and 100 nm. The calculation of steady state heat flux from  fluctuational electrodynamics principles (indicated as `FE') from Eq. \ref{QFD} is also shown. 
 Subplots b), c) and d) show the contribution of heat flux as a function of the wave-vector component $|\mathbf{k}|$ in units of Wm$^{-1}$  for time instants  $t\Gamma = 0.5, 2$ and 6 respectively for gap $d=5$ nm, and compared with the wavevector contribution to the steady state heat flux. }
\label{FigMain}
\end{figure}

Consider first the steady state expectation value of the heat exchanged due to coupling of two oscillators in contact with heat baths as given by Eq. \ref{eqHO} in the long time limit. This can be easily shown to reduce to:
  \begin{equation}
 \langle P \rangle_{t\rightarrow \infty} = \hbar \omega_{0} \dfrac{ g^2 \gamma n_1}{\gamma^2 + g^2}
 \label{eqHOSteady}
 \end{equation}
%
 The corresponding steady state expectation value of heat flux between the two half-spaces due to coupling of surface modes is obtained from employing Eq. \ref{eqHOSteady} in Eq. \ref{Q},  making the substitutions as denoted in Eq. \ref{gamma} and  Eq. \ref{gvarywithgap}. Taking $|\mathbf{k}|d = x$  the resulting expression for the steady state heat flux between two half surfaces reduces to:
\begin{equation}
 P_{\text{st}}(d) = \frac{\hbar \Gamma }{4\pi d^2} \int_0^\infty  \frac{\omega_0^{'3}(x) \xi^{'2}(x) }{\Gamma^2 +  \omega_0^{'2}(x) \xi^{'2}(x)} \frac{1}{e^{\hbar \omega_0'(x)/(k_B T)} - 1} x \, dx
 \label{eqSteady}
 \end{equation}

\begin{figure}[h!]
\includegraphics[scale=0.35]{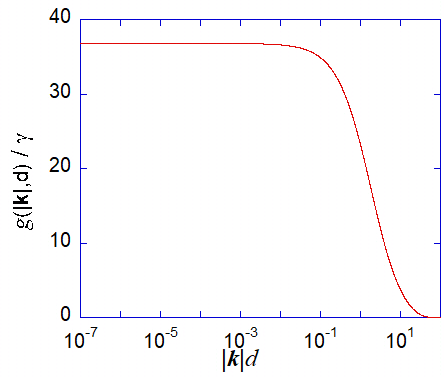}
\caption{Plot of $g(\mathbf{k},d)/\gamma$  as a function of $|\mathbf{k}|$  at $d=5$ nm}
\label{Figg}
\end{figure}

For demonstrating the dynamics of heat transfer we take the particular case of two SiC half spaces separated by a vacuum gap.  We use the following parameters for SiC \cite{palik1}:  $\epsilon_\infty =6.7$, $\omega_L = 969$ cm$^{-1}$;  $\omega_T  = 793$ cm$^{-1}$ and  $\Gamma = 4.76$ cm$^{-1}$ from which we obtain the surface polariton frequency for a single half-space $\omega_0\approx 948$ cm$^{-1}$. The master equation approach can thus be used to describe dynamic evolution of the system for time scales $t \gg 30$  fs. We choose SiC as the material of our half-spaces as the surface polariton frequency falls in the infrared frequency spectrum at which it can be excited thermally at room temperature  (as evidenced from the peak of blackbody spectrum at 300 K).
In Fig. \ref{FigMain}a we plot the heat flux from Eq. \ref{Q} as a function of the non-dimensional time $t \Gamma$ at 300 K and for vacuum gap $d$ = 5 nm and 100 nm.
The values of heat transfer from Eq. \ref{Q} can be compared with the well-known expression for the steady-state heat transfer derived using fluctuation-dissipation theorem, $P_{\text{FE}}$, which, in the limit $|\mathbf{k}|d \rightarrow 0$, reads \cite{loomis94}:
\begin{equation}
P_{\text{FE}}(d)= \frac{\hbar}{\pi^2 d^2} \int_0^\infty \omega \, d\omega \, n_1(\omega,T)\int_0^\infty \frac{ \left(\text{Im} \varepsilon(\omega)\right)^2 x e^{-x}  \, dx}{|(\varepsilon(\omega)+1)^2 - (\varepsilon(\omega) -1)^2 e^{-x}|^2}
\label{QFD}
\end{equation}
As seen in Fig.  \ref{FigMain}a the heat transfer rate from Eq. \ref{Q} conforms to the value predicted from Eq. \ref{QFD} for time intervals $ t \Gamma \gtrapprox 5$ (see appendix for analytical proof), and for smaller time intervals $t\Gamma \approx 0.1$ the heat flux is observed to reach as high as $\approx 1.7$ times the steady state value. The heat transfer contribution as a function of in-plane wavevector component $|\mathbf{k}|$ at different time intervals $t\Gamma = 0.5, \, 2,$ and 6 are also shown in Fig. \ref{FigMain}b, \ref{FigMain}c and \ref{FigMain}d  respectively and compared with the steady state distribution obtained from using Eq. \ref{eqHOSteady} in Eq. \ref{Q}. From Fig. \ref{FigMain}b and \ref{FigMain}c a transient oscillatory behavior in the heat flux contribution is observed which is indicative of the presence of strong-coupling between the surface polaritons, and which can be confirmed from the plot in Fig. \ref{Figg} where 
values of the ratio $g(\mathbf{k},d)/\gamma > 1$ indicates the strong-coupling regime. However, the total heat flux $P$ in Fig. \ref{FigMain}a is positive for all time scales due to the positive bias along the temperature gradient.

\begin{figure}[h!]
\includegraphics[scale=0.35]{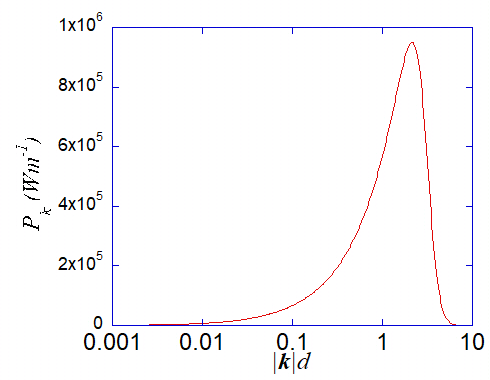}
\caption{Plot of the integrand in Eq. \ref{Q} when $t \rightarrow \infty $  as a function of $|\mathbf{k}|d$  in units of Wm$^{-1}$K$^{-1}$ at $d=5$ nm}
\label{FigCoupling}
\end{figure}

An approximate value of the nondimensional coupling parameter $\xi = 2g/\omega_{0}$,  relevant for heat transfer between two SiC half spaces can be gauged from the plot of the steady state value of the integrand in Eq. \ref{Q}  as a function of $x=|\mathbf{k}|d$  shown in Fig. \ref{FigCoupling}.
The peak value at $|\mathbf{k}|d  \approx 2$ suggests, from Eq. \ref{xip} and \ref{gamma},  that values of $\xi$  relevant for coupling of two surface modes across two SiC half spaces is approximately $\xi  \approx 10^{-2}$. In addition, the value of $\gamma/\omega_0 \approx 0.006$ validates  our assumption of weak interaction of the oscillators with the heat bath, and the subsequent use of the present form of the master equation to model the dynamic effects in heat transfer \cite{walls1970higher}. 
%

\begin{figure}[h!]
\begin{center}$
\begin{array}{cc}
\includegraphics[scale=0.35]{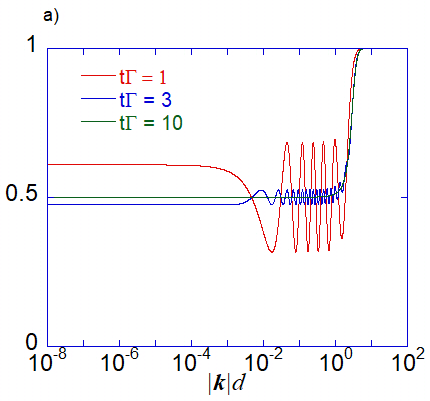}
\includegraphics[scale=0.35]{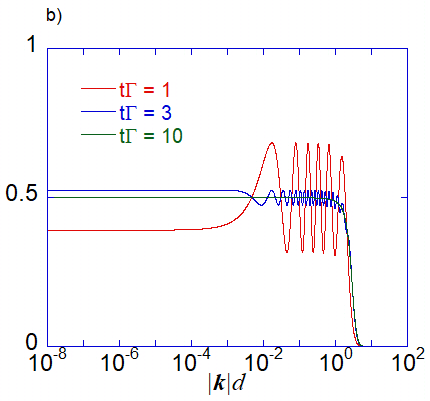}
\end{array}$
\end{center}
\caption{ Plot of a) $\langle  a^{\dagger} a \rangle/n_1$ and b) $\langle  b^{\dagger} b \rangle/n_1 $ as a function of the wavevector component $|\mathbf{k}| $ (in $m^{-1}$) at different time instants $t\Gamma = 1, 3, 10$ for $d$ = 5 nm. At time instant  $t\Gamma = 10$ steady state has been reached.}
\label{occup_k}
\end{figure}

 For large wavevectors  $|\mathbf{k}| >  1/d$ the coupling  between the oscillators drops rapidly as seen in Fig.  \ref{Figg} so that the change in thermal excitation  will be negligible at such large wavevectors.  This is also observed in Fig. \ref{occup_k}(a) and (b) where a plot of $\langle  a^{\dagger} a\rangle/n_1$ and $\langle  b^{\dagger }b \rangle/n_1$  shows  negligible change in the occupation numbers from the initial state   at these large wavevectors when steady state is reached. 
\begin{figure}[h!]
\begin{center}$
\begin{array}{cc}
\includegraphics[scale=0.35]{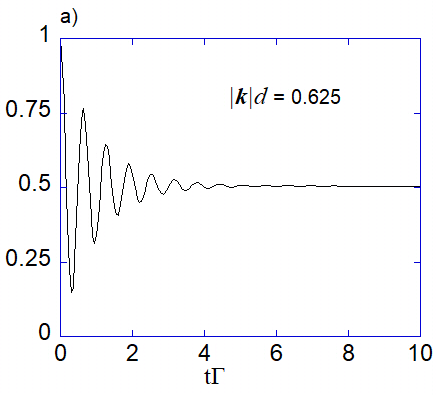}
\includegraphics[scale=0.35]{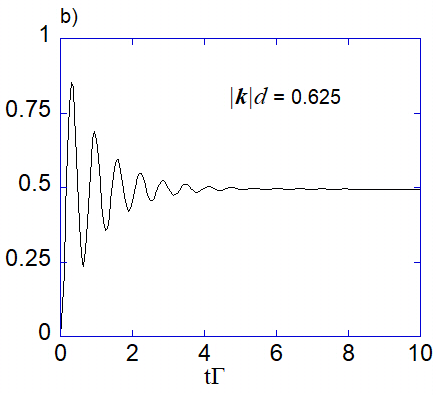}
\end{array}$
\end{center}
\caption{a) Plot of a) $\langle a^{\dagger} a \rangle/n_1$ and b) $\langle b^{\dagger} b \rangle/n_1 $ as a function of time $t\Gamma$ for the modes with wave vector component $|\mathbf{k}|d = 0.625$ }
\label{occup_t}
\end{figure}
 In addition, in Fig. \ref{occup_t} the occupation numbers for the two harmonic oscillators are plotted as a function of time.  It is observed that both oscillators attain the same steady state occupation number within a time interval $\approx 10 t \Gamma$, with the temperature difference between the heat reservoirs providing the gradient for continued heat exchange  in the steady state.

To conclude, we have shown that the theory of dynamics of coupled harmonic oscillators connected to heat baths can be used to quantify the contribution of surface modes to the dynamic near-field heat transfer between two half-spaces separated by a vacuum gap. For two SiC half-spaces it is observed that steady-state is reached for time scales $t\Gamma \gtrapprox 5$ (which corresponds to $t \approx$ 50 picoseconds) and for smaller time scales heat flux can be as high as 1.7 times the steady state value.  Experimental verification of these results would require not just the ability to measure heat transfer between macroscopic objects for small spacings but also fast response time with picosecond resolution. Recent experimental advancements show that it is  possible to measure heat transfer values for gaps as low as 0.2 - 7 nm between a STM tip and a flat substrate \cite{kloppstech2017giant}, and sub 100 nm for that between flat surfaces \cite{song2016radiative}. These measurement techniques would have to be combined with ultrafast thermometry methods (such as transient thermoreflectance technique \cite{caffrey2005thin}) in order to be able to measure the dynamic values of near-field heat transfer between macroscopic objects.  
Since, with advancement in nanotechnology near-field heat transfer between objects has increasing significance - the heat transfer between components of a magnetic storage device which are spaced few nanometers apart is a case in point \cite{challener2009heat} -  we hope that our results, along with the other recent articles \cite{yu2017ultrafast, principi2017super},  will pave the way for experimental verification of dynamic near-field heat transfer between objects. While the description here has been provided for planar surfaces, possibility of extension of this theory to other macroscopic surfaces such as microspheres and STM tips used in near-field heat transfer measurements \cite{rousseau2009radiative, shen2009surface, kim2015radiative, kloppstech2017giant} can also be explored.

\section{Appendix}
Here, we show the equivalence of the expression of steady-state heat transfer arrived using the coupled harmonic oscillator method as given in Eq. \ref{eqSteady} with that derived from fluctuational electrodynamics principles - Eq. \ref{QFD}.  This has been shown numerically in Fig. \ref{FigMain}  for the  Lorentzian form of the dielectric function given in Eq. \ref{dielectric}.  However, the complicated expressions for the natural frequency $\omega'(|\mathbf{k}|d)$ and the coupling constant $\xi'(|\mathbf{k}|d)$, as given in Eq. \ref{omegap} and \ref{xip} respectively, preclude us from showing the analytical equivalence of these two expressions for the general form of the dielectric function. For simplicity we consider the particular case of $\varepsilon_\infty = 1$ and $\omega_T= 0$ where the expressions for $\omega'(|\mathbf{k}|d)$ and $\xi'(|\mathbf{k}|d)$ reduce to the simple forms: $\omega'(|\mathbf{k}|d) = \omega_L/\sqrt{2}$ and $\xi'(|\mathbf{k}|d) = e^{- |\mathbf{k}|d }$. Such a form of dielectric function is valid for some materials such as gold (for which the corresponding parameters are: $\varepsilon_\infty = 1$, $\omega_T= 0$,  $\omega_L= 1.71 \times 10^{16}$ s$^{-1}$, $\Gamma =  1.22\times 10^{14}$ s$^{-1}$) \cite{chapuis2008effects}.  We also assume that the temperature $T$ of the half-space to be such that $k_B T \gg \hbar \omega_0$. In these limits Eq. \ref{QFD} reduces to:
\begin{equation}
P_{\text{FE}}(d)=  \frac{k_B T}{\pi^2 d^2} \int_0^\infty x e^{-x} \, dx \int_0^\infty \frac{M(\omega)}{N(\omega)} \, d\omega
\label{cauchy}
\end{equation}
where, $M(\omega) = \Gamma^2 \omega_L^2 \omega^2  $ and $N(\omega) = 16 \Gamma^4 \omega^4 - \left(4 \omega^4 - 4 \omega_L^2 \omega^2 + (1 - e^{-x}) \omega_L^4 \right)^2 + 8 \Gamma^2 \omega^2  \left(4 \omega^4 - 4 \omega_L^2 \omega^2 + (1 + e^{-x}) \omega_L^4 \right) $.  Since the rational function $M(\omega)/N(\omega)$ (which is an even function in $\omega$) has no poles on the real axis, the integral over $\omega$ can be carried out in the complex plane using Cauchy's residue theorem. 
Equation \ref{cauchy} then reduces to:
\begin{equation}
P_{\text{FE}}(d)=  \frac{k_B T\, \Gamma}{16 \pi d^2} \int_0^\infty \frac{ \omega_L^2 x e^{-x}}{\omega_L^2 e^{-x} + 2 \Gamma^2} \, dx  ;
\label{cauchyL}
\end{equation}
By making the substitutions $\omega'(|\mathbf{k}|d) = \omega_L/\sqrt{2}$,  $\xi'(|\mathbf{k}|d) = e^{- |\mathbf{k}|d }$  and $ x = 2 |\mathbf{k}|d$ in Eq. \ref{eqSteady} the expression for $P_{\text{FE}}(d)$ in Eq. \ref{cauchyL} can be shown to match that for  $ P_{\text{st}}(d)$ in Eq.  \ref{eqSteady}.  A similar derivation for showing this equivalence can be found in Ref. \cite{barton2015classical}.
\section*{Acknowledgements}
This project has received funding from the European Union's Horizon 2020 research and innovation programme under the Marie Sklodowska-Curie grant agreement No 702525.  We would like to acknowledge useful discussion with Dr. Svend-Age Biehs.  The collaboration was strengthened by a two week visit of K.S. to the Texas A\&M University.

\section*{References}


\end{document}